# ENABLING TOMORROW'S PLANETARY DEFENCE AND SPACE RESOURCE ECONOMY
## Autonomous fleet-based asteroid rendezvous missions

Stefania Soldini[1,*], Paul A. Abell[2], Daniel J. Scheeres[3], Yuichi Tsuda[4], Xiaoyu Fu[1], Nicolò Stronati[1], Hanjoon Shim[1], Sui Chen[1], Anirudh Chhabra[1], Ricardo Torres[1] and Andrew Jones[1]

[1]Department of Mechanical and Aerospace Engineering, University of Liverpool, UK

[2]Chief Scientist for Small Body Exploration, NASA, USA

[3]Colorado Center for Astrodynamics, University of Colorado, USA

[4]Institute of Space and Astronautical Science (ISAS), JAXA, Japan

*Contact email: UKRI-FL Fellow and Senior Lecturer in Space Engineering (stefania.soldini@liverpool.ac.uk)

*A White Paper for the UK Space Agency 2035 Space Frontiers programme*

**Thematic Area:** *The Solar System & Planetary Environments (Planetary)*

Call for White Papers - GOV.UK

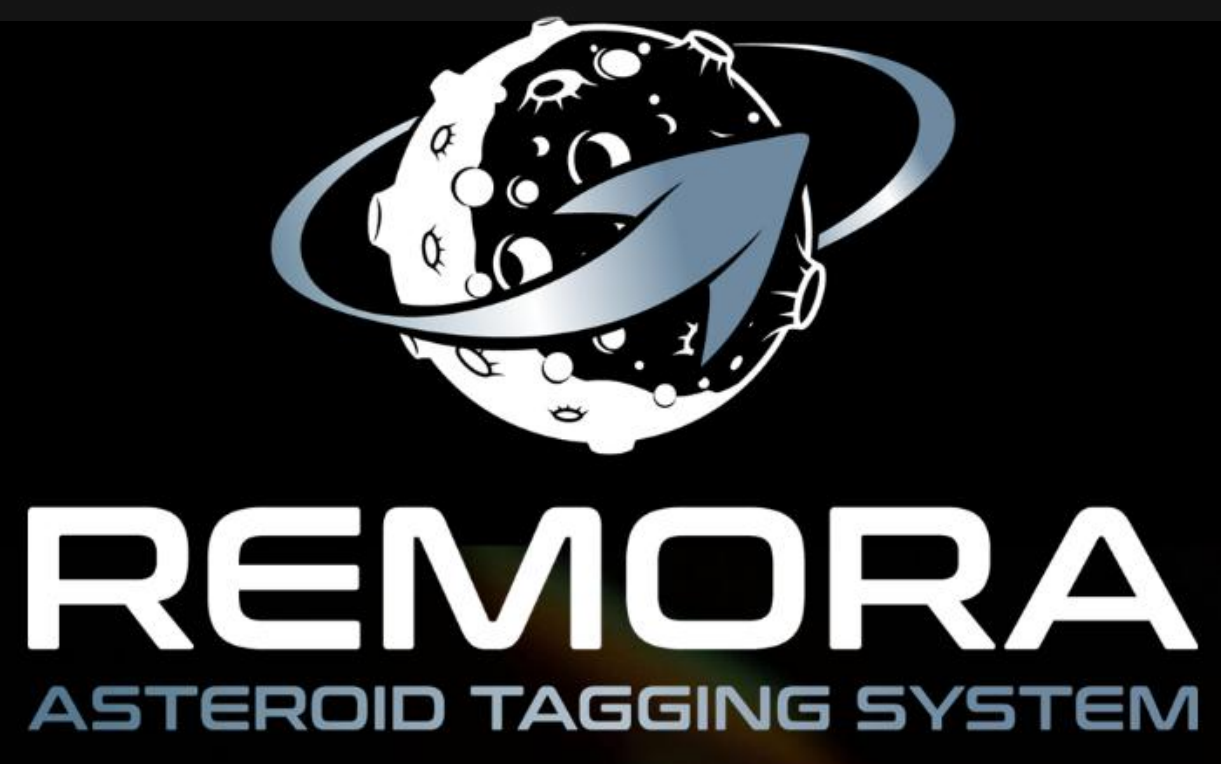

**Contents**

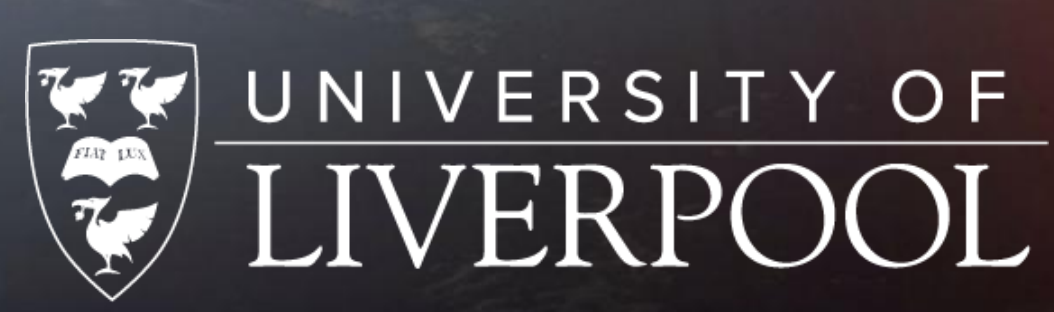

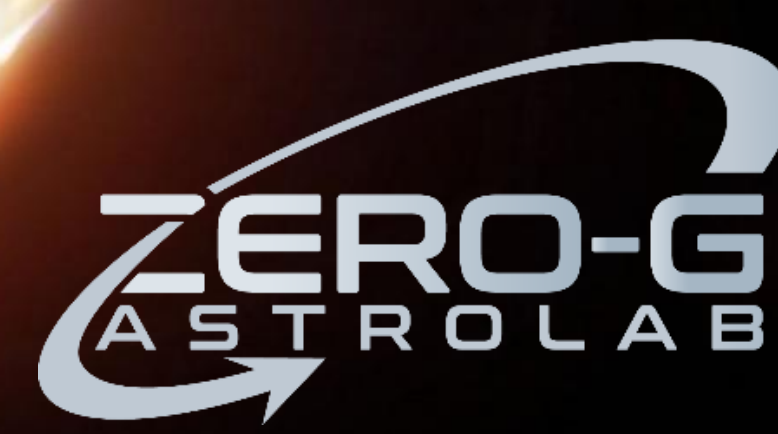

**SUMMARY.** Asteroids preserve the solar system's earliest history and pose real threats to Earth. Strengthening the UK's capabilities to detect, track, and characterise Near-Earth Objects (NEOs) is vital for national security, world-leading planetary science, and future space resource opportunities. The UK has been an influential contributor to planetary defence, from establishing the UK NEO Task Force in 2000 [1] to active roles in the **International Asteroid Warning Network (IAWN)** [2], the **Space Mission Planning Advisory Group (SMPAG)** [3], and the development of the **National Space Operations Centre (NSpOC)** [4]. UK scientists contribute to major international asteroid missions including NASA's OSIRIS-REx, DART, Lucy, and Psyche; ESA's Hera and RAMSES; and JAXA's Hayabusa2 and MMX. However, the UK currently lacks dedicated funding streams to deliver asteroid missions. Ground-based observations of asteroids can't definitively determine the physical characteristics of these objects, which are crucial for impact-risk assessment, deflection strategy, or resource evaluation.

This white paper proposes UK leadership in autonomous, low-cost asteroid-rendezvous missions, leveraging technologies developed through the UKRI-funded REMORA programme [5]. We outline four priorities:

1. strengthen NEO detection capabilities;
2. reinforce UK participation in international planetary defence missions;
3. develop autonomous rendezvous and in-situ characterisation technologies; and
4. enable the future space resource economy through targeted asteroid exploration.

Together, these actions position the UK to lead rapid, affordable deep-space missions and secure long-term strategic advantage.

**SCIENTIFIC MOTIVATION.** Asteroids and comets preserve the primitive materials from which the Earth formed 4.5 billion years ago, and collisions with Near-Earth Objects (NEOs) have shaped our planet's evolution. **Characterising NEOs is therefore essential both for understanding our solar system and for safeguarding the UK population, economy, and space infrastructure.** The UK has a strong heritage in planetary science and has led major advances in solar system formation research. The 2022 Solar System Advisory Panel (SSAP) roadmap identifies a key priority: ***"P1.5 What are the internal and surface properties of comets and asteroids, how did they form and evolve, and how do they interact with their environment?"*** [6].
Asteroid composition holds key information on early solar-system evolution, Earth's formative processes, and the ingredients that support habitability [7]. The SSAP roadmap emphasises the importance of **in-situ surface investigations** and improved **Space Situational Awareness (SSA)** to protect UK space assets and enhance the UK's contribution to planetary defence.

The UK has played a pivotal role in shaping global planetary defence policy. The 2000 UK Task Force on NEOs [1] was one of the first national efforts to assess NEO risk and helped inform later international initiatives, including ESA's Near-Earth Object Coordination Centre (NEOCC, 2013) [8] and NASA's Planetary Defense Coordination Office (PDCO, 2016) [9]. The UK remains active through the National Space Operations Centre (NSpOC, 2024) [4], IAWN [2], and SMPAG [3]. However, planetary defence is still not fully integrated into the UK's national space safety priorities, which currently emphasise space weather and debris.

As of November 2025, **40,099** Near-Earth Asteroids (**NEAs**) are known, including **2,513** larger than 140 m that qualify as Potentially Hazardous Asteroids (**PHAs**) [10]. The long-term trajectories of these small, inactive bodies (1 m-100 km) are difficult to predict due to gravitational perturbations, meaning precise astrometric tracking is essential, particularly for objects like Apophis. Ground-based systems also suffer from significant detection blind spots, especially for objects approaching from the Sun's direction, as demonstrated by the Chelyabinsk impact in 2013 [11] and recent discoveries such as asteroid 2024 YR4 (~60 m) [12,13]. Early observations of Apophis (~350 m) in 2004 indicated a possible 2029 impact, prompting major international monitoring efforts. Although that threat has since been ruled out, the Chelyabinsk event, causing **~1,600 injuries without warning**, showed that **even 20 m-class objects** can pose significant risk when detected too late.

These events raise a fundamental question:

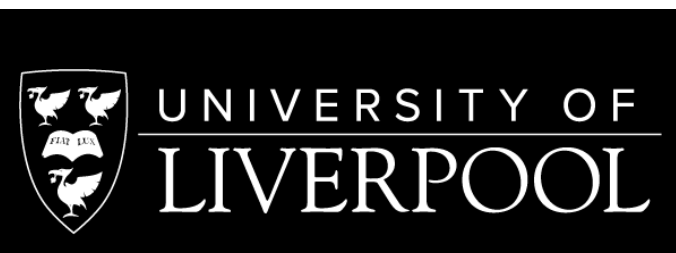

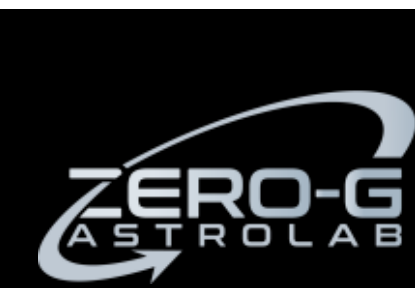

***How can the next generation of space-based technology complement ground-based observations to systematically characterise Asteroids at higher levels of detail?***

Ground-based observations alone can't precisely determine an asteroid's physical characteristics or internal mass distribution with sufficient accuracy for impact-risk assessment or deflection mission design [1]. Since an asteroid's response to any mitigation technique depends on its composition, porosity, and internal cohesion, spacecraft rendezvous missions remain essential for enabling robust dynamical modelling and asteroid's physical characterisation. Despite decades of observations and missions, our understanding of asteroid interiors is still limited [14], largely because mass estimates alone can't resolve internal density variations [15].

Recognising this gap, the UK's Near-Earth Objects Task Force Report (2000) [2] called for **affordable rendezvous missions using microsatellites** to characterise different asteroid types. The UKRI-funded **REMORA** project (2022–present) [5] directly advances this national vision. REMORA is developing a scalable architecture in which autonomous CubeSat beacons rendezvous with multiple asteroids, enabling precise orbital reconstruction and characterisation. This approach supports future planetary defence operations, enhances planetary science return, and provides the foundational data needed to assess potential asteroid resources.

Beyond planetary defence, asteroids hold potential as **resources for future space activities**. Many are rich in ***water, a key element for producing rocket propellant supporting both human and robotic exploration and for enabling in-space construction and reducing costs in spaceflight***. Indeed, water is essential for life support, radiation shielding for astronaut safety and in-space manufacturing of rocket fuel for oxygen and hydrogen. ***Asteroids are often rich in valuable raw materials*** (e.g., platinum group metals), that could be used for material processing. Difficulties in unlocking asteroid resources include the high cost of spaceflight, unreliable identification of asteroids' physical characteristics and their suitability for mining, and challenges in extraction. In the long-term, unlocking ***such key asteroid data could in turn enable an emerging asteroid mining market*** [16]. International interest is already evident through initiatives such as Luxembourg's **Space Resource Initiative** [17], which reflects growing economic and strategic attention to the utilisation of asteroid materials as well as the **new Japanese Center for Space Resources and Innovation** jointly operated by a few institutions in Japan [18].

Addressing these scientific and strategic challenges will require a coordinated programme of targeted research, technology development, mission innovation as well as international cooperation with the already existing initiatives, priorities that form the basis of the objectives outlined below.

**OBJECTIVES.** The UK is a global leader in planetary science, contributing to major international asteroid missions for both science and planetary defence. UK laboratories have analysed samples from Bennu, Ryugu, and Itokawa [7], and UK scientists play key roles in flagship missions including NASA's DART, OSIRIS-REx, Lucy and Psyche; ESA's Hera and RAMSES; and JAXA's Hayabusa2 and MMX. However, this participation is not supported by **dedicated domestic funding**, international agencies do not cover foreign contributions, and UKSA's general planetary science calls have not targeted asteroid missions. The UK also leads in small-satellite technology, notably through Surrey Satellite Technology Ltd (SSTL). CubeSats and deployable payloads/landers are increasingly used to enhance the capability of larger missions, as demonstrated by JAXA's Hayabusa2, NASA's DART, ESA's Hera and RAMSES companion payloads, and NASA's Mars Cube One (MarCO), highlighting the opportunity for UK-led small spacecraft to play a growing role in planetary exploration. With SSTL contributing to recent ESA Mini-F Phase 1 proposals, the UK is well placed to lead future low-cost, rapid-development asteroid missions. A UKSA-funded Phase 0 study would strengthen national mission concepts, enhance competitiveness in ESA Mini-F/F-class selections, and build a pipeline toward future bilateral missions. These strengths underpin the objectives that follow, defining the scientific and strategic advances needed to sustain UK leadership in planetary defence and small-body exploration:

*<u>O1. Improve NEO Detection and Tracking</u> [19]*: Strengthen UK capability to detect and track NEAs and PHAs by investing in national ground-based assets such as Jodrell Bank radar and the UK

Fireball Alliance (UKFAll) [20]. Address critical blind spots, especially sunward-approaching objects, through participation in ESA's NEOMIR and FlyEye initiatives and improved monitoring from L1/L2, enhancing both UK and European resilience.

*O2. Enhance UK Leadership in International Planetary Defence [19]*: Maintain and expand the UK's contributions to IAWN, SMPAG, and missions such as ESA's Hera and RAMSES. Because UK scientists currently lack dedicated domestic funding for participation in international missions (e.g., DART, OSIRIS-REx, Lucy, Hayabusa2, MMX), we recommend UKSA mechanisms enabling UK-led payloads, Mini-F/CubeSat platforms, and supported researcher time, enabling competitive UK participation in ESA/NASA/JAXA missions and strengthening national resilience to asteroid threats.

*O3. Develop Autonomous Rendezvous Capability for Asteroid Characterisation [2,14]:* Accelerate low-cost autonomous proximity-operations technology, such as distributed CubeSat systems like REMORA, to deliver detailed in-situ characterisation of asteroids beyond the limits of ground-based observations [1,13]. This will build sovereign expertise in spacecraft autonomy and deep-space operations while creating industrial opportunities in sensing, navigation, and mission services.

*O4. Enable a Future Space Resource Economy [21]:* Characterise asteroid composition to assess mining potential and in-situ resource utilisation (ISRU), including water extraction for propellant, life support, and radiation shielding. Progress in this area directly supports national commercial opportunities identified in the recent House of Lords report on space resources [21].

*This white paper focuses on Objectives 3-4 to demonstrate how UK technological strengths can be translated into leadership in the emerging space resource economy, while indirectly advancing Objectives 1-2.*

**STRATEGIC CONTEXT.** UK Space economy and small satellite industry: The miniaturisation of spacecraft technologies now enables CubeSats to operate in deep space, as demonstrated by missions such as ASI's LICIACube, ESA's Juventas and Milani, and NASA's MarCO, CAPSTONE, Janus and M-Argo. The UK is exceptionally well positioned to lead this evolution, with one of the world's most advanced small-satellite industries, recognised for innovation, agility, and cost-effective mission delivery. SSTL, a global pioneer in small-satellite engineering, has delivered more than 70 spacecraft and supported key ESA programmes including Galileo, Lunar Pathfinder, and Mini-F concepts. The UK also benefits from a rapidly growing ecosystem of SMEs and major space contractors, including Open Cosmos, In-Space Missions, AAC Clyde Space, Teledyne e2v, Oxford Space Systems, MDA UK, Thales Alenia Space UK, and Airbus Defence & Space UK, providing end-to-end capability across avionics, propulsion, payloads, deployable structures, and operations.

The UK space sector contributes over **£17.5B** to the economy, supports **55,000+ jobs**, and generates **£5.8B in exports** [22]. With new launch infrastructure in Scotland and Cornwall, the UK is becoming a launch-capable nation, further enabling rapid deployment of small satellites into deep-space trajectories. Together, this industrial base, scientific excellence, and launch capability form a coherent national ecosystem capable of delivering low-cost, rapid-development missions, including autonomous asteroid rendezvous, planetary defence demonstrators, and deep-space science, on timelines under five years, a critical requirement for effective planetary defence.

Planetary defence: Asteroid missions have been driven primarily by NASA (USA), ESA (Europe), and JAXA (Japan), with China now emerging as a major actor through new sample-return and planned deflection missions. Additional nations, including India, South Korea, and the UAE, are entering the field, notably with the UAE's **Mission to the Asteroid Belt** (**MBR) Explorer** (launch 2028, asteroid landing). China is rapidly advancing its planetary defence capabilities: it has launched **Tianwen-2** (2025), is planning a **kinetic-impactor demonstration** (target potentially **2015 XF261**), and is developing dedicated **early-warning and monitoring networks** [23]. This expanding international participation highlights a shifting global landscape in which multiple nations are now investing in NEO detection, characterisation, and mitigation.

Planetary resources and asteroid mining: National and commercial interest in space resources is accelerating, closely aligned with long-term goals in asteroid characterisation and in-situ resource utilisation (ISRU). UKSA is already investing in lunar resource extraction, such as the **DISRUPT-2**

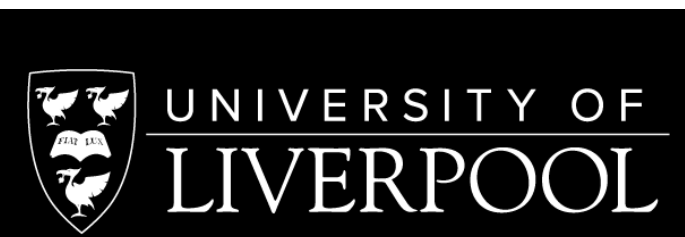

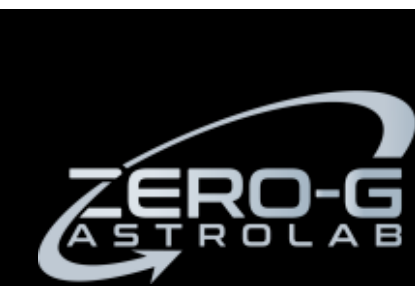

ISRU project (Thales Alenia Space, Oxfordshire), to support future human exploration, deep-space staging, and sustainable operations using locally sourced materials [24]. Although lunar-focused, these initiatives signal the UK's intent to develop space resource capability directly relevant to asteroids. In 2024, UKSA also commissioned a major **Space Resource Utilisation** study from Space Professionals Partnership (SPPL) examining regulatory and capability needs across the lunar and cislunar environment and asteroids [25]. Industry is likewise emerging in this sector, including the **Asteroid Mining Corporation**, exploring technologies for microgravity extraction and magnetic-levitation simulation. Together, these efforts show a strategic opportunity for the UK to position itself early in the developing asteroid-resource economy.

*Apophis 2029:* Apophis presents a unique opportunity for public engagement and international cooperation. Its close approach in April 2029, passing within geostationary orbit and visible from the UK, coincides with the UN's *International Year of Asteroid Awareness and Planetary Defence*. It provides a timely platform to highlight UK contributions and to collaborate with the United Nations Office for Outer Space Affairs (UNOOSA) on establishing a Panel on Asteroid Orbit Alteration (PAOA) to ensure responsible governance of future deflection activities. With rising global activity, including China's planned impact-deflection test in 2028, the UK is well placed to lead on planetary defence policy and coordination [19].

**PROPOSED APPROACH.** The REMORA Project: REMORA (REndezvous Mission for Orbital Reconstruction of Asteroids) is a UKRI-funded programme developing a new mission architecture in which a fleet of self-driven CubeSats autonomously rendezvous with and characterise multiple asteroids. The concept is inspired by the symbiotic relationship between remora fish and sharks: small, agile CubeSats ("remoras") attach or station-keep/land near larger bodies ("sharks", i.e., asteroids) to obtain continuous, high-precision orbital and physical measurements, Fig. 1.

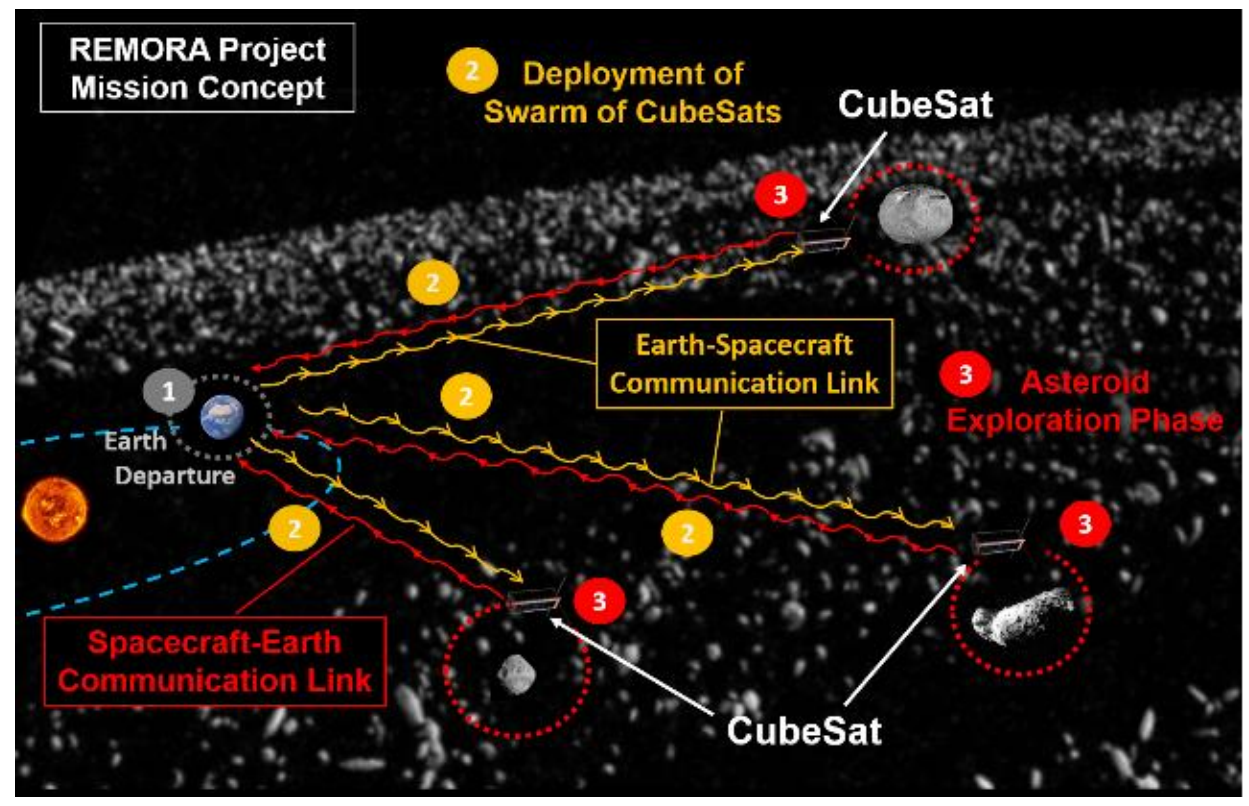

***Figure 1**. REMORA Project Mission Concept.*

REMORA is structured as a 4+3-year programme:

- *Phase 1 (Years 1-4, end Jan 2028)* develops autonomous navigation and path-planning software for operations around diverse asteroid types and gravitational environments.
- *Phase 2 (Years 5-7, ~2031)* will mature this technology into a mission concept in which a fleet of six CubeSats independently rendezvous with six distinct asteroids.

*In the near-term*, REMORA will deliver NEAR (Near-Earth Asteroid Regions), an onboard path-planning tool enabling fuel-minimal manoeuvring at reduced operational cost. NEAR will be validated using a new microgravity hardware-in-the-loop (HIL) testbed at the University of Liverpool's Zero-G AstroLab, alongside the group's research on inferring asteroid interior structure from the dynamical response of a spacecraft.

*Our long-term vision* is that REMORA provides the foundation for a low-cost, scalable, multi-mission deep-space architecture, enabling equitable access to planetary science and supporting:

1. Planetary defence, through improved tracking and characterisation of PHAs;
2. Planetary science, by advancing understanding of asteroid formation and solar-system evolution; and
3. Space resource utilisation, by generating the data needed to identify and evaluate asteroid resources.

University of Liverpool Contributions and Capability: The University of Liverpool's Zero-G AstroLab, Fig. 2, led by Dr Stefania Soldini, is advancing methods for physical characterisation based on spacecraft dynamics and developing a dedicated HIL facility to demonstrate autonomy in proximity operations. The Zero-G AstroLab conducts research in astrodynamics, guidance, navigation and control (GNC), spacecraft autonomy and proximity operations, and machine-learning techniques for gravity inversion and interior reconstruction, supported by an ESA-OSIP grant (PI: Dr Soldini)
Research Focus Areas: The group is advancing three core technical areas. First, interior-modelling and gravity-inversion methods integrate ground-based observations with spacecraft measurements using Machine Learning-based approaches, developed through ongoing ESA-OSIP and UKRI-REMORA projects. Second, the team models asteroid dynamical environments by computing periodic-orbit families and characterising near-surface dynamics, providing insights into regolith mobility, surface processes, and spacecraft-asteroid interactions (e.g., published models for Ryugu: Fu, X., Soldini. S [26-27] and Song, Z., Yu, Y., Soldini, S. et al. [28]). Third, ongoing research on manoeuvre optimisation and autonomous path planning develops scalable navigation and trajectory-design methods for multi-target missions, including techniques for autonomous landing, gravity-field measurement and in-situ operations (e.g. published Hayabusa2 mission operations: Soldini et al. [29-31]). These algorithms will be validated using the physical HIL testbed, enabling realistic demonstrations of GNC autonomy under representative proximity-operations conditions.

**PROPOSED TECHNICAL SOLUTION & REQUIRED DEVELOPMENT:** The University of Liverpool is developing the core technologies required for autonomous, low-cost asteroid rendezvous missions. Through the UKRI-funded REMORA programme, we are building a generalised framework that enables CubeSats to operate safely and efficiently around a wide range of asteroid types. We are advancing three integrated technical capabilities for scalable asteroid rendezvous. 1) dynNEAR develops generalised dynamical models for spacecraft motion around diverse asteroid types, monolithic, contact-binary, and binary systems, across sizes, compositions, and rotation states, providing a unified basis for mission design. 2) goNEAR applies Dynamical Systems Theory to identify natural orbits and fuel-efficient pathways, enabling predictable, low-cost CubeSat operations in irregular gravity fields. 3) autoNavNEAR delivers autonomous navigation and path-planning, combining advanced GNC and ML-supported decision-making to ensure safe, fuel-minimal proximity operations around small bodies. These algorithms are being validated in the **new Zero-G AstroLab** at the **University of Liverpool** (Fig. 2), a **high-precision 2.5 × 5 m microgravity test facility** featuring the flattest floor in the UK and one of its kind in Europe, a precision epoxy air-bearing system, and a Qualisys motion-capture network for high-fidelity hardware-in-the-loop demonstrations of autonomous asteroid-rendezvous and proximity-operations technologies. Together, these developments create a scalable technical foundation for future UK-led Mini-F missions, multi-target CubeSat fleets, and autonomous deep-space exploration capabilities. The facility is also designed to support a broad range of space technology applications, including active space debris removal, distributed spacecraft and swarm concepts, docking, and deployable/origami space structures.

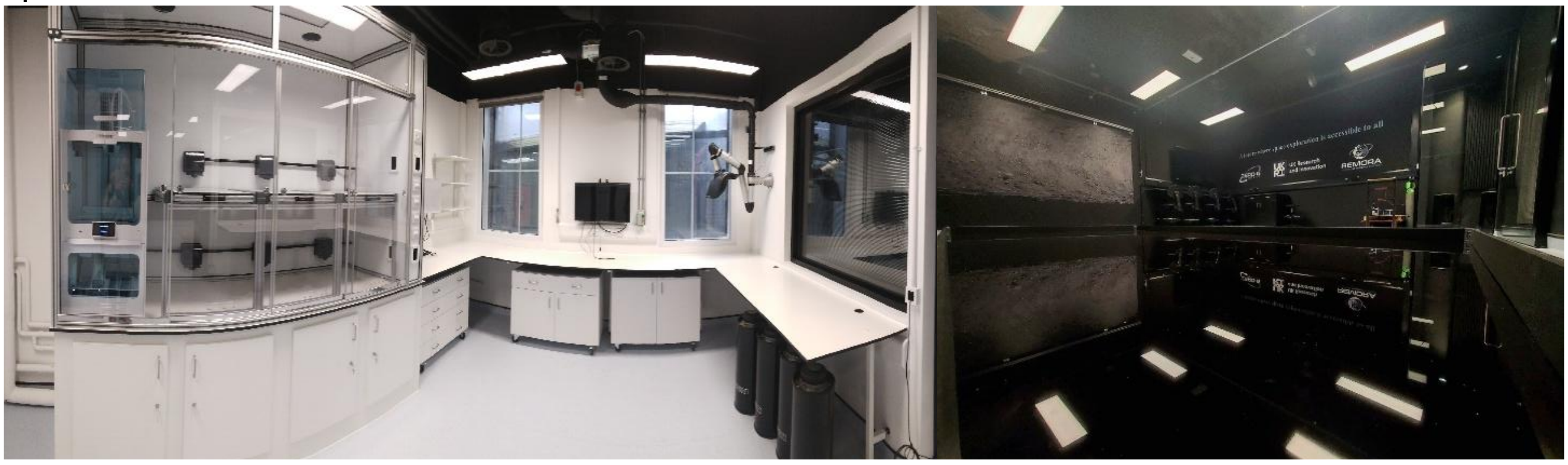

***Figure 2**. University of Liverpool's Zero-G AstroLab microgravity test facility, featuring a 2.5 m × 5 m Precision Epoxy air-bearing floor (mean flatness 0.27 mm; typical point-wise deviation ~0.01 mm; maximum inclination 0.11 mm over 5 m, equivalent to ~0.022 mm/m) and a Qualisys motion-capture tracking system for high-fidelity hardware-in-the-loop proximity-operations testing.*

**UK LEADERSHIP & CAPABILITY:** The UK has a long and distinguished record in planetary defence and small-body exploration. Following the landmark 1994 Shoemaker-Levy 9 impact, the UK recognised the need for coordinated international action and, in 2000, established one of the world's first national NEO task forces [1]. Its recommendations pre-dated and informed later institutional efforts such as ESA's Planetary Defence Office, demonstrating early UK strategic leadership.

Today, the UK remains an active contributor to global coordination frameworks: it is a founding member of the UN-endorsed Space Mission Planning Advisory Group (SMPAG) and, since 2024, a member of the International Asteroid Warning Network (IAWN). The establishment of the National Space Operations Centre (NSpOC) in 2024 further strengthens UK national resilience, while the UKSA regularly convenes the national NEO community to support these international responsibilities.

UK scientists have a long history of contributing to international missions to asteroids and comets. This includes major roles in NASA's OSIRIS-REx (Bennu sample return), DART (planetary defence impact), and Lucy missions; ESA's Hera, Rosetta, and Giotto missions; and JAXA's Hayabusa and Hayabusa2 sample-return programmes, with UK laboratories analysing returned samples from asteroids Bennu, Ryugu, and Itokawa. UK research groups have provided leadership in instrument development, mission operations, ejecta and impact modelling [32-37], surface characterisation, and small-body dynamics. The UK is also contributing to ESA's emerging RAMSES mission concept to Apophis and plays an active role in future missions such as JAXA's MMX. This sustained engagement highlights the UK's depth of expertise and its long-term commitment to small-body exploration and planetary defence. The University of Liverpool's Zero-G AstroLab is actively contributing autonomy, ejecta and dynamical-modelling expertise [33-37] to major international missions including NASA's DART, ESA's Hera and RAMSES, and JAXA's Hayabusa2-SHARP.

**PARTNERSHIP OPPORTUNITIES:** ESA, NASA (e.g., OSIRIS-REx, NEAR Shoemaker), and JAXA demonstrate successful models for building national capability in small-body exploration, JAXA through fully home-grown asteroid missions (Hayabusa, Hayabusa2), and agencies such as Agenzia Spaziale Italiana (ASI) and Centre National d'études Spatiales (CNES) through highly competitive CubeSat and lander contributions to international asteroid missions (e.g., LICIACube, MASCOT). These programmes have enabled their countries' industrial sectors to flourish, with companies such as Tyvak International, GomSpace, and Argotec becoming globally recognised providers of deep-space-capable small satellites and specialised payloads.

The UK could follow a similar trajectory by developing sovereign capability while also delivering high-value CubeSat, Mini-F, and payload contributions to major ESA, NASA, and JAXA asteroid missions, strengthening both national science return and the competitiveness of the UK space sector. This strategy aligns closely with the recommendations of the US Planetary Science Decadal Survey, which calls for expanded planetary defence missions and greater international collaboration, reinforcing the importance of UK participation in global efforts. Investing now in these partnerships will maximise UK scientific return, strengthen national security, and grow the competitiveness of the UK space sector in the fast-emerging deep-space economy [38].

**SUGGESTED MISSION CLASS:** The UKRI-funded REMORA project provides a strong foundation for developing a UK-led Mini-F class deep-space mission, similar to those currently pioneered by SSTL but tailored for planetary science and small-body exploration. Mini-F missions are defined by a Cost at Completion (CaC) below €50M, a development timeline of under five years, and an emphasis on increasing programme diversity, training the next generation, expanding launch opportunities, and enabling innovative implementation schemes.

SSTL has already played a key role in recent ESA Mini-F Phase 1 proposals, and early collaboration between UK industry and academia will be essential to maximise competitiveness.

Initiating a UKSA-funded Phase 0 study would significantly strengthen mission concepts, improving prospects for ESA Mini-F/F class submissions and for pursuing subsequent bilateral mission phases. Such early alignment ensures science priorities, engineering feasibility, and affordability remain tightly integrated, critical for delivering rapid, cost-effective missions capable of progressing from concept to launch within five years. *By investing now, the UK can secure sovereign capability in autonomous deep-space operations, strengthen global planetary defence, and lead the next era of deep space exploration and space resource innovation.*

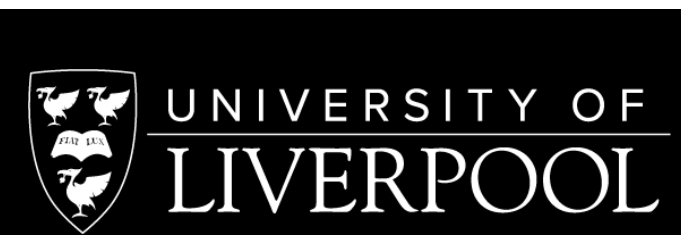

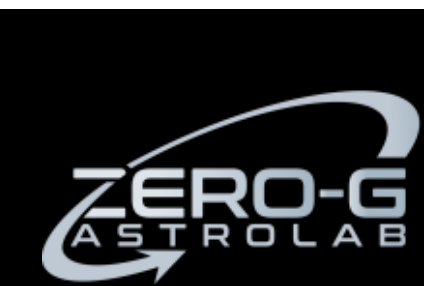

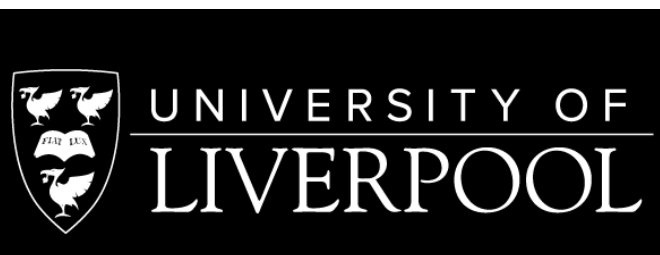

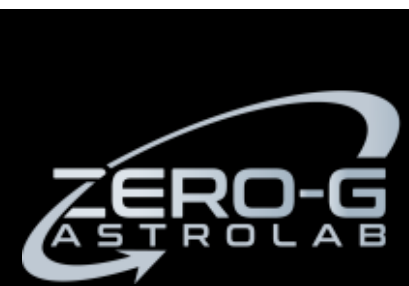

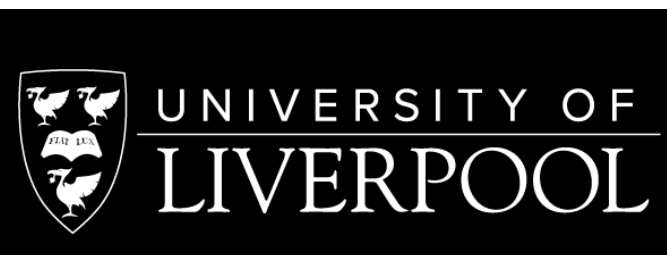

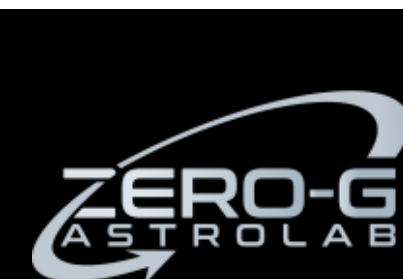

**List of Signatories**

Mark Burchell, University of Kent (UK)

Davies Faye, The Open University (UK)

Ashley King, Natural History Museum (UK)

Stephen Loury, UK Comet Interceptor group (UK)

Aleksandra Sokolowska, UKRI Fellow, Imperial College London (UK)

*REMORA Project Advisory Board:*

Paul A. Abell, NASA (USA)

Elisabet Canalias, CNES (France)

Ian Carnelli, ESA (France)

Daniel Jones, UK Space Agency (UK)

Michael Küppers, ESA (Spain)

Bob Morris, The Northern Space Consortium (UK)

Aurélie Moussi, CNES (France)

Paolo, Paoletti, University of Liverpool (UK)

Ettore Perozzi, ASI (Italy) – retired

Daniel Scheeres, University of Colorado (USA)

Martin Sweeting, Surrey satellite Technology Ltd (SSTL) (UK)

Yuichi Tsuda, Institute of Space and astronautical Science (ISAS), JAXA (JAPAN)

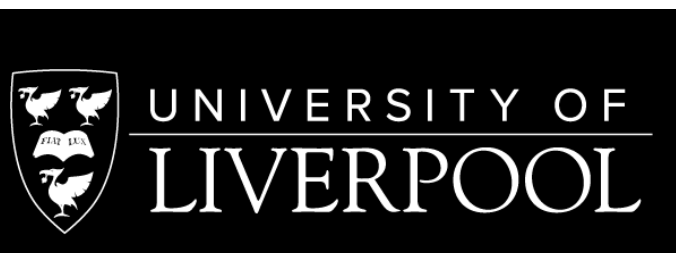

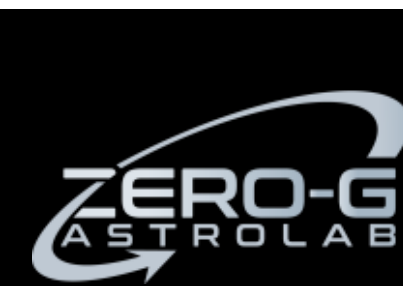